\documentclass[pra,twocolumn,superscriptaddress,twocolumn]{revtex4-1}
\usepackage[cp1251]{inputenc}
\usepackage{textcomp}
\usepackage{graphicx}
\usepackage{graphics}
\usepackage{amsmath}
\usepackage{amssymb}
\usepackage{amsfonts}
\usepackage{color}
\usepackage{amsfonts}
\usepackage{textcomp}
\usepackage[normalem]{ulem}
\usepackage{placeins}
\usepackage{tabularx}
\usepackage{booktabs}
\usepackage{textcomp}

\usepackage{hyperref}

\begin{document}
\title{Novel magnetic stoichiometric superconductor  EuRbFe$_4$As$_4$}
\author{
T.~K.~Kim
}
\affiliation{Diamond Light Source, Harwell Campus, OX11 0DE Didcot,  United Kingdom}
\author{K.~S.~Pervakov}
\author{V.~A.~Vlasenko}
\author{A.~V.~Sadakov}
\author{A.~S.~Usoltsev}
\affiliation{P.~N.~Lebedev Physical Institute, Russian Academy of Sciences, 119991, Moscow, Russia}
\author{D.~V.~Evtushinsky}
\affiliation{Laboratory for Quantum Magnetism, Institute of Physics,
\'{E}cole Polytechnique F\'{e}d\'{e}rale de Lausanne, CH-1015 Lausanne, Switzerland}
\author{S.~W.~Jung}
\affiliation{Diamond Light Source, Harwell Campus,  OX11 0DE Didcot, United Kingdom}
\author{G.~Poelchen}
\affiliation{Institut f\"{u}r Festk\"{o}rper- und Materialphysik, Technische Universit\"{a}t Dresden, D-01062, Dresden, Germany}
\affiliation{European Synchrotron Radiation Facility, 71 Avenue des Martyrs, Grenoble, France}
\author{K.~Kummer}
\affiliation{European Synchrotron Radiation Facility, 71 Avenue des Martyrs, Grenoble, France}
\author{D.~Roditchev}
\affiliation{LPEM, ESPCI Paris, PSL Research University, CNRS, 75005 Paris, France}
\affiliation{Sorbonne Universite, CNRS, LPEM, 75005, Paris, France}
\affiliation{Moscow Institute of Physics and Technology, 141700 Dolgoprudny, Russia}
\author{V.~S.~Stolyarov}
\affiliation{Moscow Institute of Physics and Technology, 141700 Dolgoprudny, Russia}
\affiliation{N.~L.~Dukhov All-Russia Research Institute of Automatics, 127055 Moscow, Russia}
\affiliation{National University of Science and Technology MISIS, 119049 Moscow,  Russia}
\author{I.~A.~Golovchanskiy}
\affiliation{Moscow Institute of Physics and Technology, 141700 Dolgoprudny, Russia}
\affiliation{N.~L.~Dukhov All-Russia Research Institute of Automatics, 127055 Moscow, Russia}
\affiliation{National University of Science and Technology MISIS, 119049 Moscow,   Russia}
\author{D.~V.~Vyalikh}
\affiliation{Donostia International Physics Center, 20018 Donostia-San Sebasti\'{a}n, Basque Country, Spain}
\affiliation{IKERBASQUE, Basque Foundation for Science, 48013 Bilbao, Spain}
\author{V.~Borisov}
\affiliation{Institut f\"{u}r Theoretische Physik, Goethe-Universit\"{a}t Frankfurt, D-60438 Frankfurt am Main, Germany}
\author{R.~Valent\'{i}}
\affiliation{Institut f\"{u}r Theoretische Physik, Goethe-Universit\"{a}t Frankfurt,
 D-60438 Frankfurt am Main, Germany}
\author{A.~Ernst}
\affiliation{Institut f\"{u}r Theoretische Physik, Johannes Kepler Universit\"{a}t, A 4040 Linz, Austria}
\affiliation{Max-Planck-Institut f\"{u}r Mikrostrukturphysik, D-06120 Halle, Germany}
\author{S.~V.~Eremeev}
\affiliation{Institute of Strength Physics and Materials Science, Russian Academy of Sciences, 634055 Tomsk, Russia}
\author{E.~V.~Chulkov}
\affiliation{Donostia International Physics Center, 20018 Donostia-San Sebasti\'{a}n, Basque Country, Spain}
\affiliation{Saint Petersburg State University, 198504 Saint Petersburg, Russia}
\affiliation{Departamento de F\'{i}sica de Materiales UPV/EHU and Centro de F\'{i}sica de Materiales (CFM-MPC),
Centro Mixto CSIC-UPV/EHU, 20080 Donostia-San Sebasti\'{a}n, Basque Country, Spain}
\author{V.~M.~Pudalov}
\affiliation{P.~N.~Lebedev Physical Institute, Russian Academy of Sciences, 119991, Moscow, Russia}

\maketitle

Superconducting compounds, including atoms with a large spin magnetic moment (Gd$^{3+}$, Eu$^{2+}$, etc.),
are remarkable objects for studying such fundamental problem of condensed matter physics
as the interplay of magnetic ordering and superconducting pairing \cite{zapf_RPP_2017}.
Magnetism has traditionally been viewed as a phenomenon antagonistic towards superconductivity; nevertheless
as is now well known, there are many classes of materials with unconventional superconductivity,
from heavy fermion compounds, borocarbides, to copper oxides, in which the superconducting state is
closely related or
even arises from magnetism.

Among recently intensively studied iron pnictides, the EuFe$_2$As$_2 $  compound
\cite{paglione_NatPhys_2010, si_NatRevMat_2016} ignited a live interest in that  beside
the AFM-ordering of  Fe-electrons in a spin-density-wave type structure  (at high temperatures $\sim 195$K),
at lower temperatures $\sim 19$\,K, in the superconducting region, a long-range magnetic order
sets in the
Eu$^{2+}$. In the low-temperature magnetic state, the spins of the electrons of Eu are ordered ferromagnetically in
the  $ab$ plane  and antiferromagnetically along the  $c$ axis \cite{dutta_JPCM_2013}.
As a result, the original parent compound EuFe$_2$As$_2 $ is an insulator, but when doped or
when the lattice is deformed under pressure, it becomes a superconductor.

Thus, in doped EuFe$_2$(As$_{1-x}$P$_x$)$_2$, for $x \sim 0.2$, superconductivity below
 $T_m \approx 19$\,K coexists with  magnetically-ordered subsystem of Eu electrons
 \cite{ren_PRL_2009, cao_JPCM_2011, nowik_JPCM_2011, stolyarov_SciAdv_2018, veshchunov_JETPL_2017,
 nandi_prb_2014, devizorova_PRB_2019, devizorova_PRL_2019, ghigo_PRR_2019, grebenchuk_PRB_2020}.
Such coexistence does not lead to deviations of the superconducting state from the conventional
one with the singlet  type pairing \cite{ren_PRL_2009, cao_JPCM_2011}.
This contrasts with the known behavior in borocarbides, for which the lattice doping with rare-earth magnetic
ions leads to a decrease in the critical temperature
  \cite{muller_RPP_2001}.
The apparent complete independence of the superconducting properties of electrons from the magnetic
 ordering of the Eu sublattice is
associated with the multiorbital nature of iron-based superconductors,
in which the exchange interaction and superconductivity are provided by
different groups of electrons.

In 2016, a new family of superconducting iron pnictides was discovered, the so-called ``1144'' type compounds
$AeA$Fe$_4$As$_4$, (where $Ae$ = Ca, Sr, La, Eu and $A$ = K, Rb, Cs) \cite{iyo_JACS_2016}. These compounds
  demonstrate more intriguing properties than ``Eu-122'', due to the fact that they are superconductors in
  stoichiometric composition.
As in ``122''-compounds, atoms with a large spin magnetic moment (Gd $^{3+}$, Eu$^{2+}$
  and others) built into the crystal lattice tend to magnetic ordering.

The Meissner state in magnetic superconductors of the ``1144'' family turned out to be nontrivial:
it coexists with short-scale ferromagnetic domains with dimensions smaller than the field penetration
depth and at low temperatures transforms into a very
interesting domain-vortex state \cite{veshchunov_JETPL_2017, stolyarov_SciAdv_2018}.
Small sizes of magnetic domains are necessary for the existence of a superconducting state at
temperatures below $T_m$ \cite{anderson_PR_1959}. In this state,
self-generated entrance of vortices at temperatures near $ T_m $  was discovered in
Ref.~\cite{vlasko-vlasov_PRB_2019}. The increase in magnetic induction $ B $ observed in
\cite{vlasko-vlasov_PRB_2019} is most pronounced at $ T_m $. This unusual behavior of $ B $
 as opposed to the usual pushing out of the magnetic field in conventional superconductors,
is a consequence of the collective reaction of the magnetic and superconducting subsystems
in EuRbFe$_4$As$_4$.

The unusual properties of the magnetic superconducting compound motivated us to carry out comprehensive studies
of the thermodynamic magnetic properties, the amplitude of the superconducting order parameter, as well as
the band structure, particularly
in the vicinity of the Fermi level for carriers responsible for superconductivity,
and the structure of the electronic states of
Eu and its spin state.

\begin{figure}
    \includegraphics[width=0.3\textwidth]{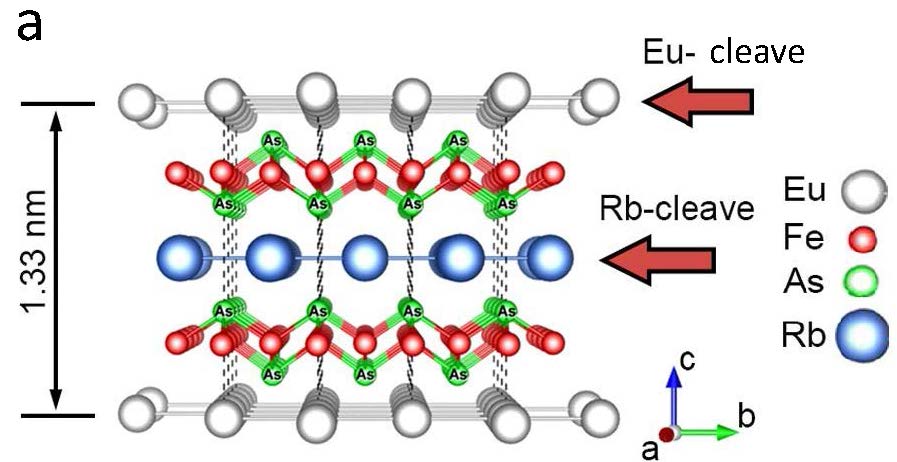}
\hspace{0.05in}
 \vspace{0.3in}
    \includegraphics[width=0.16\textwidth]{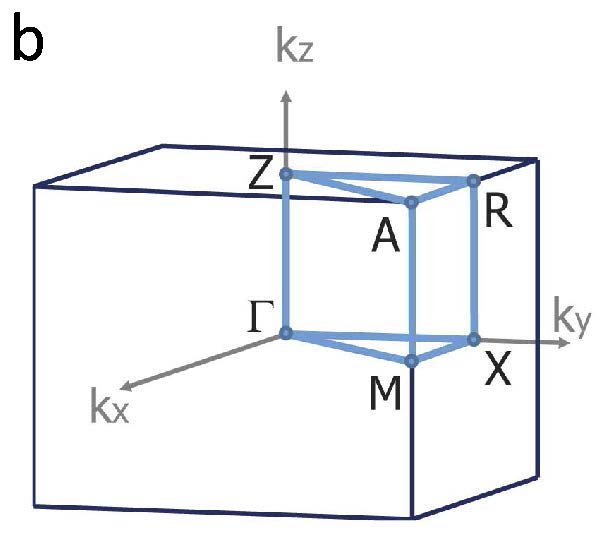}
\caption{(a) Crystal structure of EuRbFe$_4$ As$_4$. Red arrows show the
cleavage planes (Quoted from \cite{stolyarov_JPCL_2020}); (b) bulk Brillouin zone of RbEuFe$_4$As$_4$.
}
\label{fig:structure}
\end{figure}

\section{Experiment}

\subsection{Crystal lattice}
the  $AeA$Fe$_4$As$_4$ compounds have the tetragonal crystal lattice (P4/mmm), in which the $Ae$ and $A$
layers form two nonequivalent  structures \cite{iyo_JACS_2016}. In particular,
in EuRbFe$_4$As$_4$ (Fig.~\ref{fig:structure}), in contrast to the `` 122 '' parent compound,
EuFe$_2$As$_2$ and RbFe$_2$As$_2$ blocks  alternate along the $ c $ axis
\cite{bao CrGr_2018, liu_PRB_2016, jiang_ChPhysB_2013, iyo_JACS_2016,
liu_SciBull_2016}.
As a result of the different ionic radii of Eu and Rb atoms have, such a two-block structure
turns out to be strained. This factor, as well as the redistribution between
Eu- and Rb-
layers of a small amount (5-10\%) of atoms lead to self-doping of the nominally stoichiometric compounds
with $p$-type carriers  (0.25 holes/Fe atom);
as a result, a superconducting state emerges with a critical temperature $T_c \approx 36$\,K.
Magnetic ordering in the Eu sublattice
occurs at a temperature $T_m \sim 15$\,K, inside the domain of the superconducting state.
Due to  the missing
mirror symmetry along the $ c $ axis, in EuRbFe$_4$As$_4$ (in contrast to EuFe $ _2 $ As $ _2 $),
the magnetic order emerging  below the Curie temperature  is a helicoidal antiferromagnet structure
\cite{devizorova_PRB_2019, iida_PRB_2019, koshelev-helical_PRB_2019}.

\subsection{Samples}
\begin{figure}
    \includegraphics[width=0.3\textwidth]{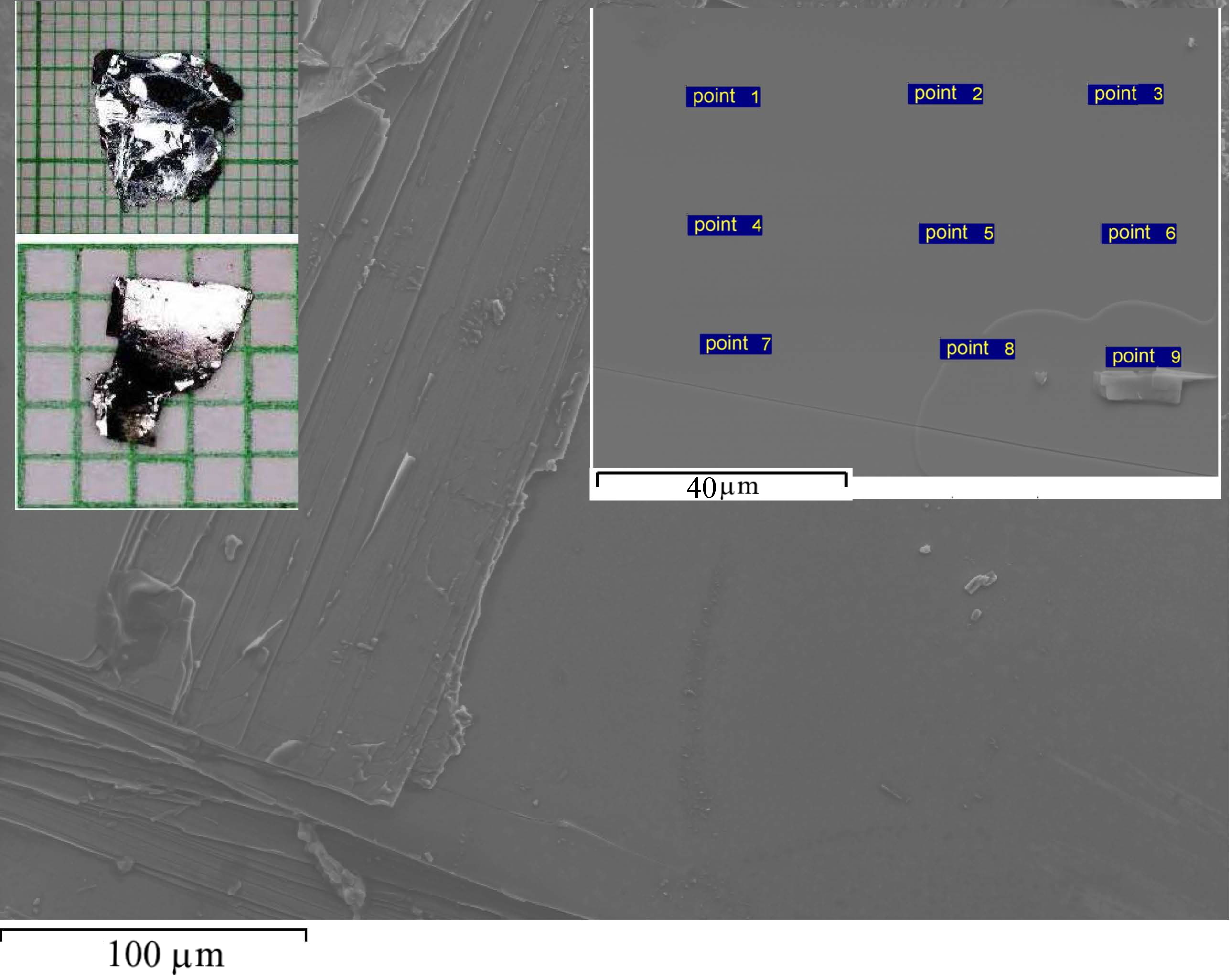}\\
\caption{SEM image of the crystal cleavage surface. Left inset: optical image; right:
enlarged image of a flat area of the surface showing the regions of the EDX measurement of
elemental composition.
}
\label{fig:EDX}
\end{figure}

High-quality single crystals of EuRbFe$_4$As$_4$ were grown by the self-flux technique.
the initial high purity components of Eu (99.95\%), Rb (99.99\%)
and the pre-synthesized precursor FeAs (99.98\% Fe + 99.9999\% As) were mixed with
1:1:12 molar ratio.  The mixture was placed into an alumina crucible and
vacuum-tightly welded into an Nb container under a  0.2\,atm residual pressure of argon.
The container was then placed in a tube oven with an argon atmosphere and the furnace
was heated to 1250$^\circ$C, kept at this temperature for 24\,h for
homogenization of the melt, and then cooled to 900$^\circ $C at a rate of 2$^\circ$C/h.
At this temperature
the container with the crystal was kept for another 24 hours to reduce the concentration of
growth defects and then finally
cooled to room temperature inside the furnace. The large-size  crystals were removed from the crucible in
an Ar atmosphere in an Ar glove box. Finally, the crystals were transported to the measuring devices
in a sealed container, with minimal time exposure to air.
The appearance of representative
crystals are shown in the inset in Fig.~\ref{fig:EDX}. They had typical dimensions up to 5\,mm in the basal
plane with large areas of an atomically-flat surface (see Fig.~\ref{fig:EDX}).

\begin{table}
\caption{Elemental composition of the  RbEuFe$_4$As$_4$ crystals (in weight  \%) at 9 surface regions, shown
in Fig.~\ref{fig:EDX}}
\begin{tabular}{|c|c|c|c|c|c|}
\hline \hline
Region & Fe & As & Rb & Eu &  Total\\ \hline
1      & 28.26 & 37.98 & 9.11 & 24.65 & 100.00\\
2      & 28.37 & 38.18 & 8.79 & 24.66 & 100.00\\
3      & 28.16 & 38.09 & 8.83 &	24.91 &	100.00\\
4      & 27.96 & 38.39 & 8.81 &	24.85 &	100.00\\
5      & 27.91 & 38.49 & 8.82 &	24.78 &	100.00\\
6      & 27.86 & 38.35 & 8.68 &	25.10 &	100.00\\
7      & 27.92 & 38.39 & 8.69 &	25.00 &	100.00\\
8      & 27.81 & 38.55 & 8.73 &	24.92 &	100.00\\
9      & 28.18 & 39.05 & 9.34 &	23.43 &	100.00\\ \hline
average &28.05 & 38.39 & 8.87 &	24.70 &	100.00\\
$\sigma$ &0.20 & 0.31 &	0.22 &	0.50 &        \\ \hline \hline
\end{tabular}
\end{table}

\begin{figure}
    \includegraphics[width=0.35\textwidth]{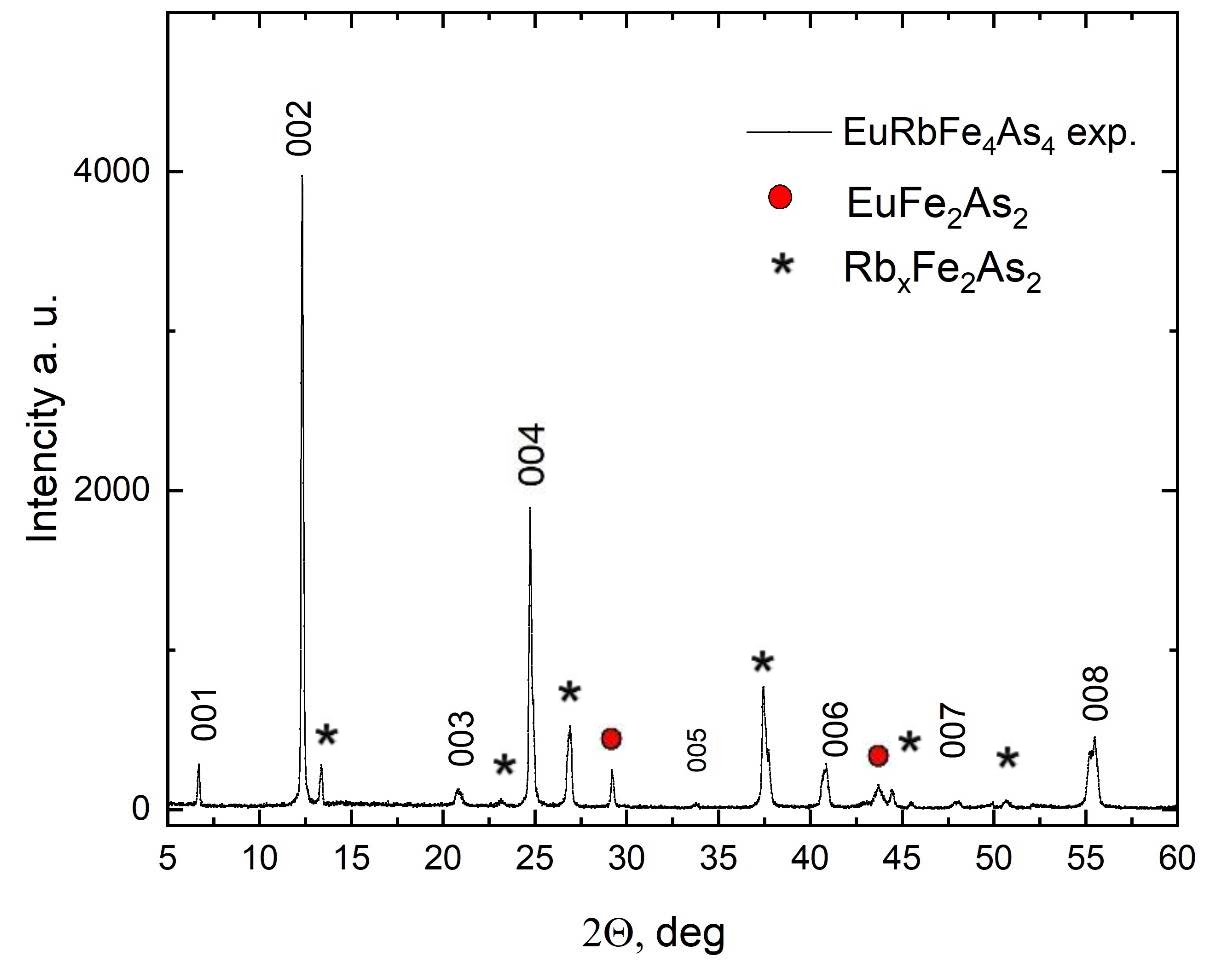}
\caption{X-ray diffraction pattern of the EuRbFe$_4$As$_4$ single crystal. The asterisk marks the peaks
corresponding to the  Eu-122 side phase. Quoted from Ref.~\cite{vlasenko_SST_2020}.
}
\label{fig:XRD}
\end{figure}

The X-ray diffraction pattern of the obtained crystals, shown in Fig.~\ref{fig:XRD}, indicates
the presence, in addition to the dominant phase of Eu-1144, a small amount of the impurity Eu-122  phase, which
inevitably arises in the process of synthesis.   Elemental analysis in the  mapping mode (Table~1)  at 9 different
regions  points
out a uniform distribution of the elements over the sample surface (see inset to Fig.~\ref{fig:EDX}).
We found
that the  factual chemical composition of our single crystal was
Eu$_{1.06(2)}$Rb$_{0.81(2)}$Fe$_{3.92(5)}$As$_{4.00(5)}$.
The minor excess of Eu atoms is due to
the presence of about 10\% EuFe2As2 phase; the TEM results reveals that this side phase
 is present as short irregulsr atomic chains between the planes of majority  Eu-1144 phase.

\subsection{Magnetic and superconducting properties from thermodynamic measurements }
\begin{figure}
    \includegraphics[width=0.35\textwidth]{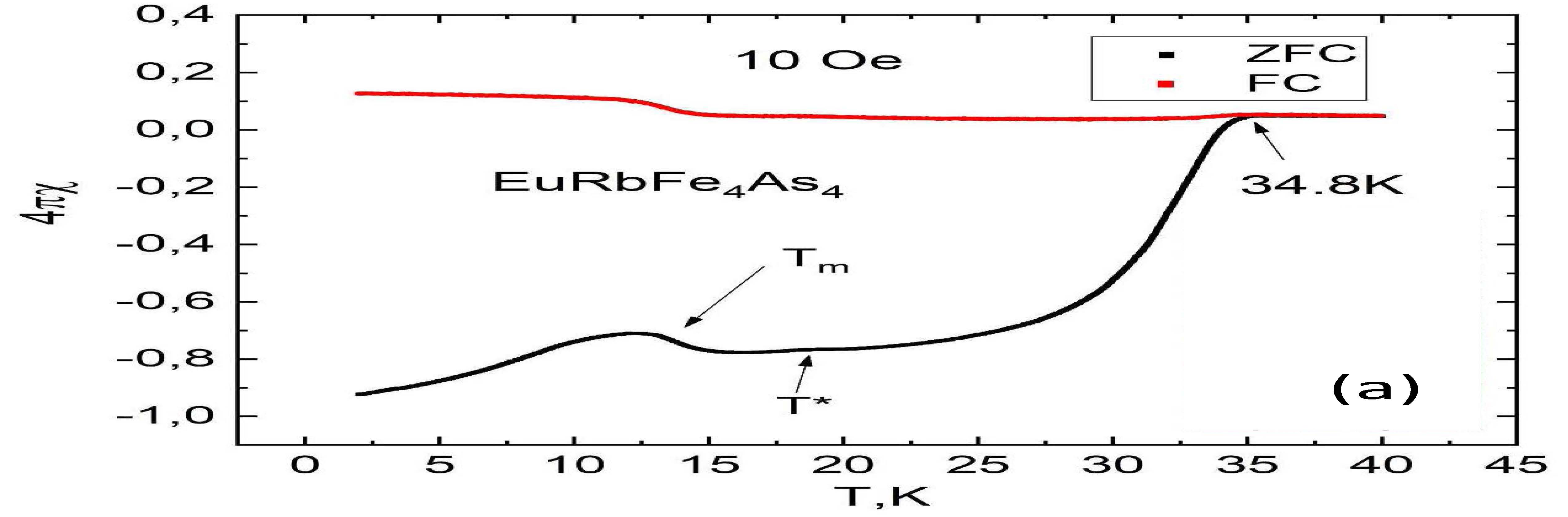}
     \includegraphics[width=0.35\textwidth]{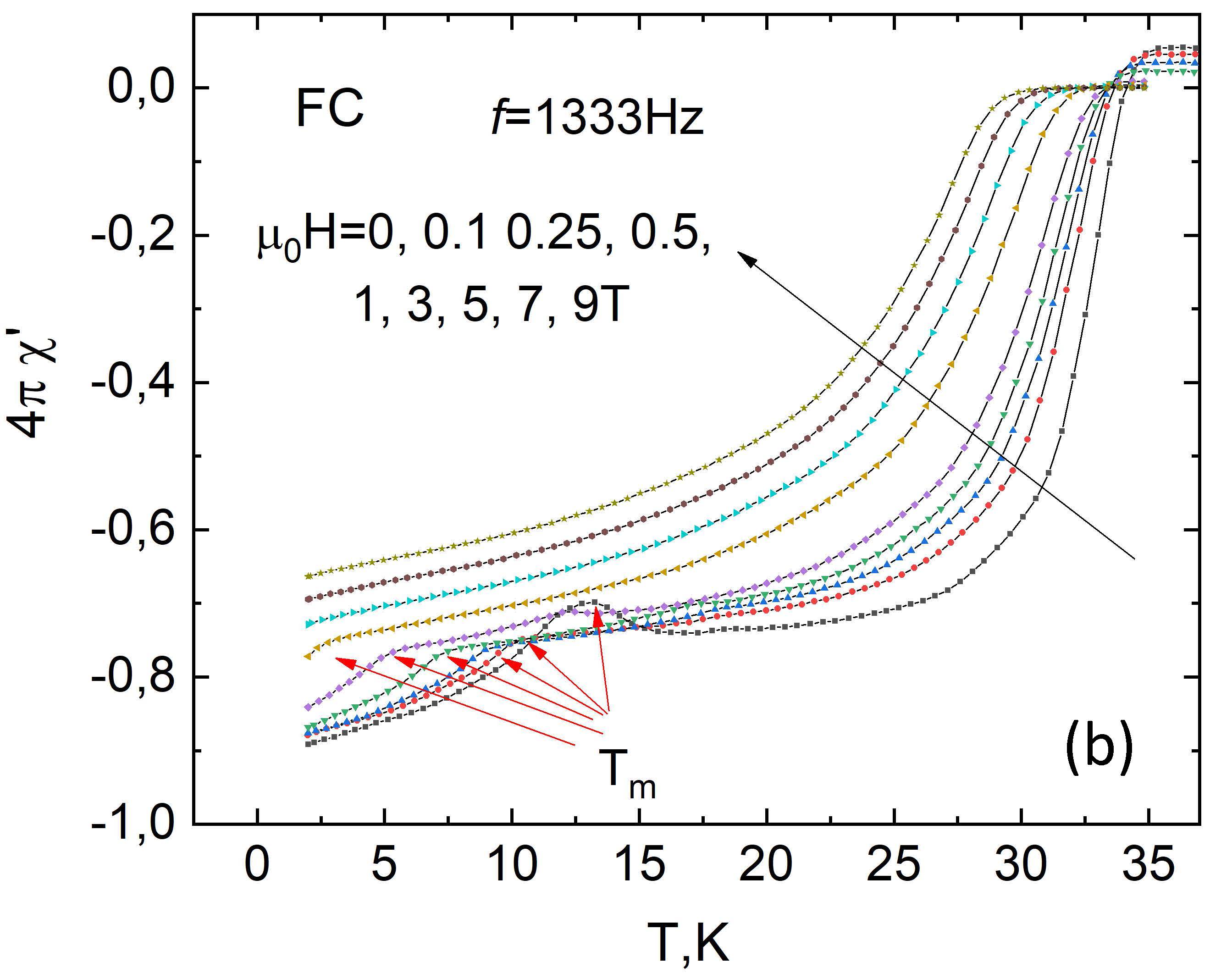}
\caption{(a) temperature dependence of DC magnetic susceptibility for cooling in zero field
(ZFC) and in an external field (FC) of 10\,Oe applied in the $ab$ plane.
(b) temperature dependence of the real part of the magnetic susceptibility at various
field values. Quoted from Ref.~\cite{vlasenko_SST_2020}.
}
\label{fig:chi}
\end{figure}

The temperature dependence of the DC magnetic susceptibility (Fig.~\ref{fig:chi}) shows a sharp diamagnetic
transition upon cooling in zero field (ZFC mode) at $T_c \approx 35$\,K due to a superconducting transition;
the volume fraction of the superconducting phase is $\sim 90$\% at 2\,K.

In the superconducting state, at $T \approx 12-16$\,K, the DC susceptibility has a peak indicating
the onset of magnetic ordering in the sublattice of Eu atoms.
The temperature dependence of the AC susceptibility (Fig.~\ref{fig:chi}) also shows a peak at the same
temperature  $T_m \approx 15$\,K. In a magnetic field $ H $ applied in the $ab$ plane, as field increases the peak
shifts towards lower temperatures and completely disappears at $ H\approx 3$\,T, which allows us
 estimate roughly the    exchange field  $\sim 3$\, T in the plane of Eu atoms. The lowest arrow in Fig.~4a marks  temperature  $T^*=19$K that is the known point of the AFM ordering in the Eu-122 compound \cite{cao_JPCM_2011}; the absence of features at 19\,K confirms that the amount of this side phase in the studied crystals is negligibly small.

\begin{figure}
    \includegraphics[width=0.35\textwidth]{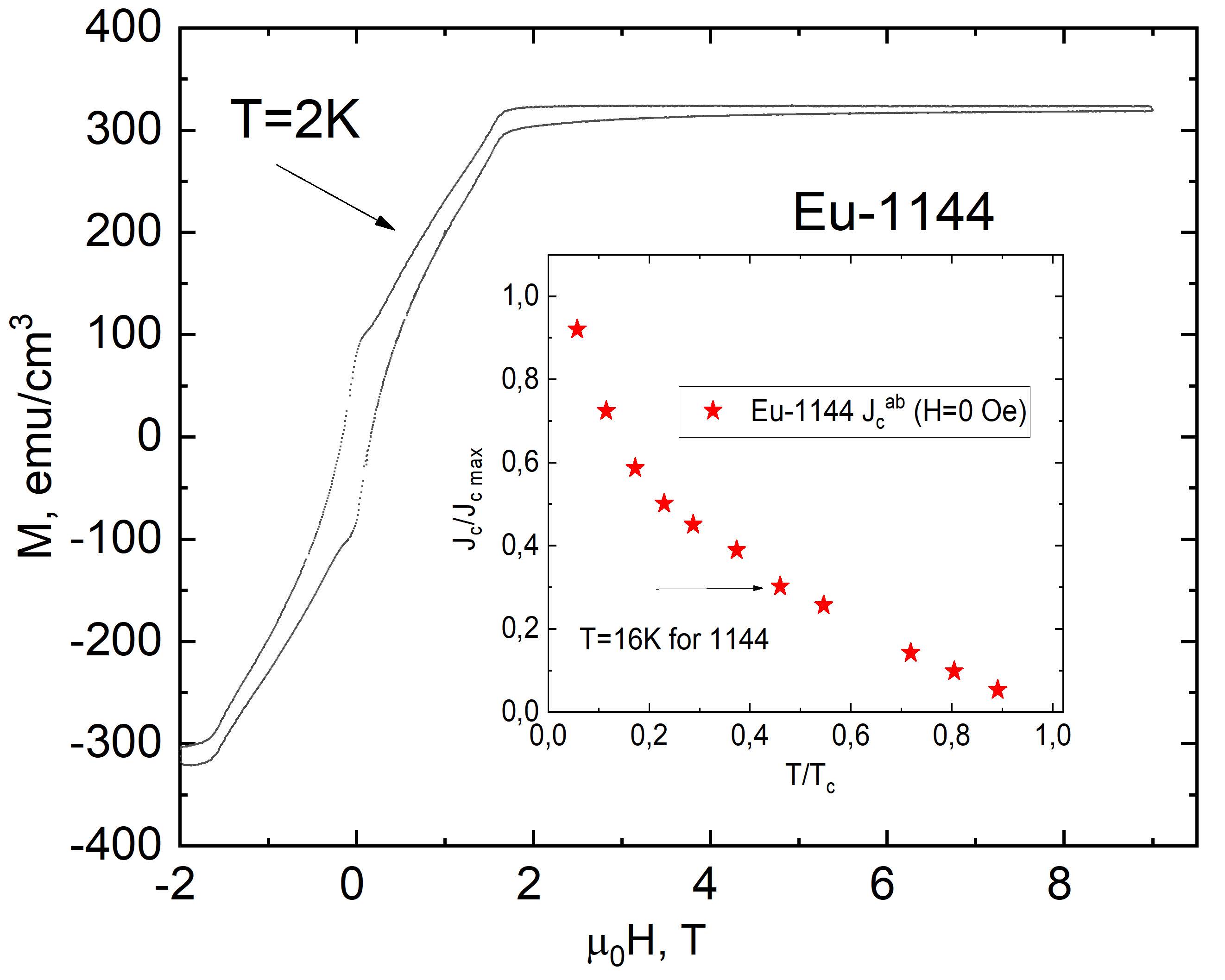}
\caption{Magnetization curve at $ T = 2$\,K. The inset shows the temperature dependence of the
critical current $ J_{ab}$, normalized to its maximum value.
Quoted from Ref.~\cite{vlasenko_SST_2020}.
}
\label{fig:M(H)}
\end{figure}

The $M(H)$ curves (Fig.~\ref{fig:M(H)}) also demonstrate the superposition of the diamagnetic
superconducting hysteresis at $ T < T_c $ associated with vortex pinning, and (at $ T <T_m$)
magnetization curves with a saturation in a field 2\,T typical for ferromagnetic ordering
\cite{vlasenko_SST_2020}.
Magnetization
$ M (H) $ measurements were carried out with field orientation in the $ ab $ plane, at a field sweep rate
100\,Oe/s. From the magnetization saturation field in Fig.~\ref{fig:M(H)} $\approx 310$\,emu/cm$^3$
at $ T = 2$\,K, we obtain an estimate for the magnetic moment of the Eu atoms $ 6.4\mu_B$, which is in
a good agreement with the tabular value $ 7\mu_B$.

The $ M$ versus  $H$   hysteresis in Fig.~\ref{fig:M(H)} practically does not change
when passing through the temperature $ T_m $; therefore, we conclude that it is associated
only with  vortices pinning in the superconducting state.
This enables us to estimate the critical current $J_c $ from the width of the hysteresis loop
$\Delta M(H) $ using
Bean models \cite{bean_RMP_1964}. The normalized  $ J_c $ values as a function  of the normalized temperature,
are shown in the inset in Fig.~\ref{fig:M(H)}. These $M(H)$ curves have a form typical for the superconducting compounds of the
``Ba-122'' and ``CaK-1144'' families, which do not contain magnetic atoms \cite{vlasenko_SST_2020}.
We  conclude that the susceptibility, high field magnetization and critical current are not sensitive
to the presence of strong magnetism of Eu atoms.

\subsection{Superconducting properties from Andreev reflection spectroscopy, ARPES and STS }
Andreev reflection spectroscopy was measured using the
break-junction technoque (BJ)
\cite{kuzmicheva_UFN_2014, kuzmicheva_EPL_2013, abdel-hafiez_PRB_2018, bezpi_PRB_2019, kuzmicheva_JETPL_2016,
NbS2_PRM_2018}.
For measurements, the sample shaped as a flat plate (in the $ab$ plane)
was attached to an isolated springy sample
stage  made of Be-bronze. The sample was fixed with a conductive
paste at 4 points to enable  4-contact measurements of the current-voltage
($I - V $) characteristics and dynamic conductance ($dI/dV $ vs $V$). To create microcrack in the bulk of the crystal,
the stage with the sample was precisely deformed in the liquid helium environment at 4.2\,K,
which prevented microcracks from contamination and oxidation. To form
the S-N-S  junction (or, more precisely, ``superconductor-constriction-superconductor''),
the microcrack edges of the crystal were procisely tuned (by mechanically bending the stage),
until the desired Andreev reflection regime was achieved. In the latter regime,  the current-voltage
characteristic  exhibits
a pronounced excess current (increase in conductivity) in the vicinity of  zero voltage
offset \cite{kummel_PRB_1990}.

In the Andreev reflection mode, within the superconducting temperature range $ T <T_c $,
on the dynamic conductivity ($ dI / dV $ vs $ V $) of the S-N-S contact
minima appear at discrete values of the bias voltage
\cite{moreland_JAP_1985, kuzmicheva_UFN_2014, kummel_PRB_1990}.
These minima are associated with the first-order resonances at $ V = 2\Delta_i/e $, where $ \Delta_i $
is the gap value for the $i $-th SC condensate, and $ e $ is the elementary charge.
When the sample is split in BJ experiments, in addition to a single S-N-S junction,
chains of series-connected S-N-S junctions often form.
For such chains, the resonant features of Andreev reflection  are observed at bias voltages
$ V_m = m\times (2\Delta_i/e)$ (where $ m = 2 $, 3, ...); the resulting spectrum may be easily dientangled
by scaling the bias voltage with respect to integer values of $ m $
\cite{kuzmicheva_EPL_2013, kuzmicheva_UFN_2014}.

\begin{figure}[h]
   \includegraphics[width=0.35\textwidth]{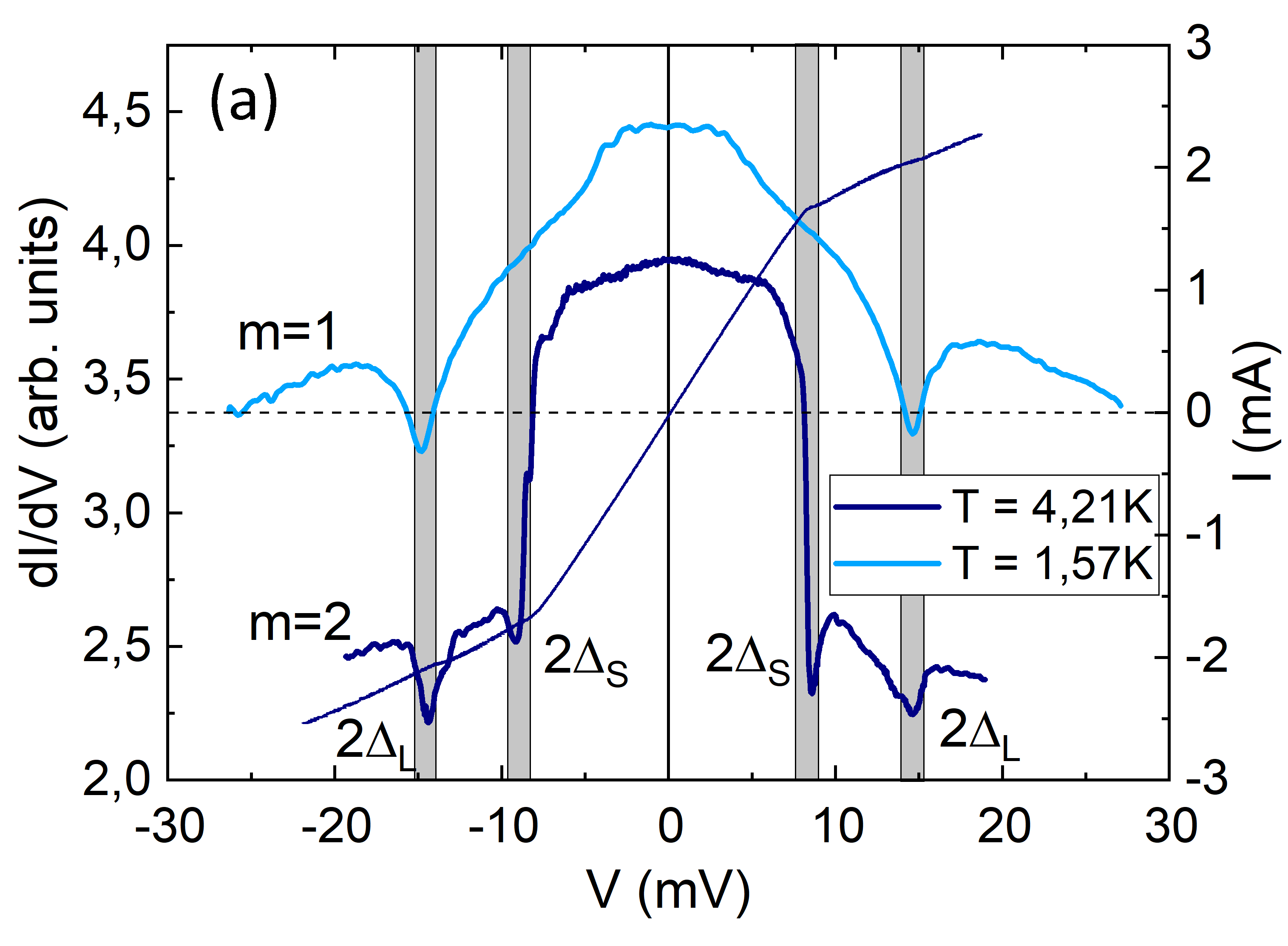}
   \includegraphics[width=0.35\textwidth]{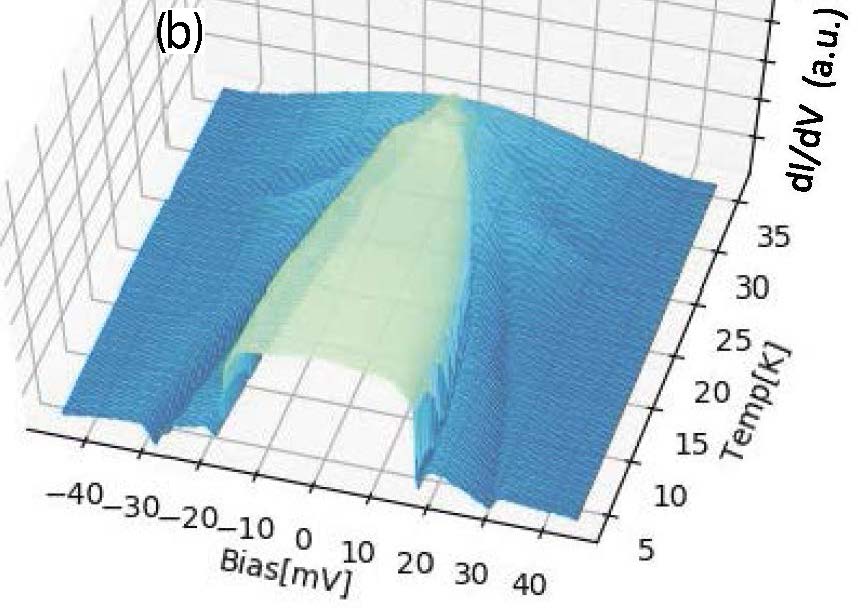}
      \includegraphics[width=0.35\textwidth]{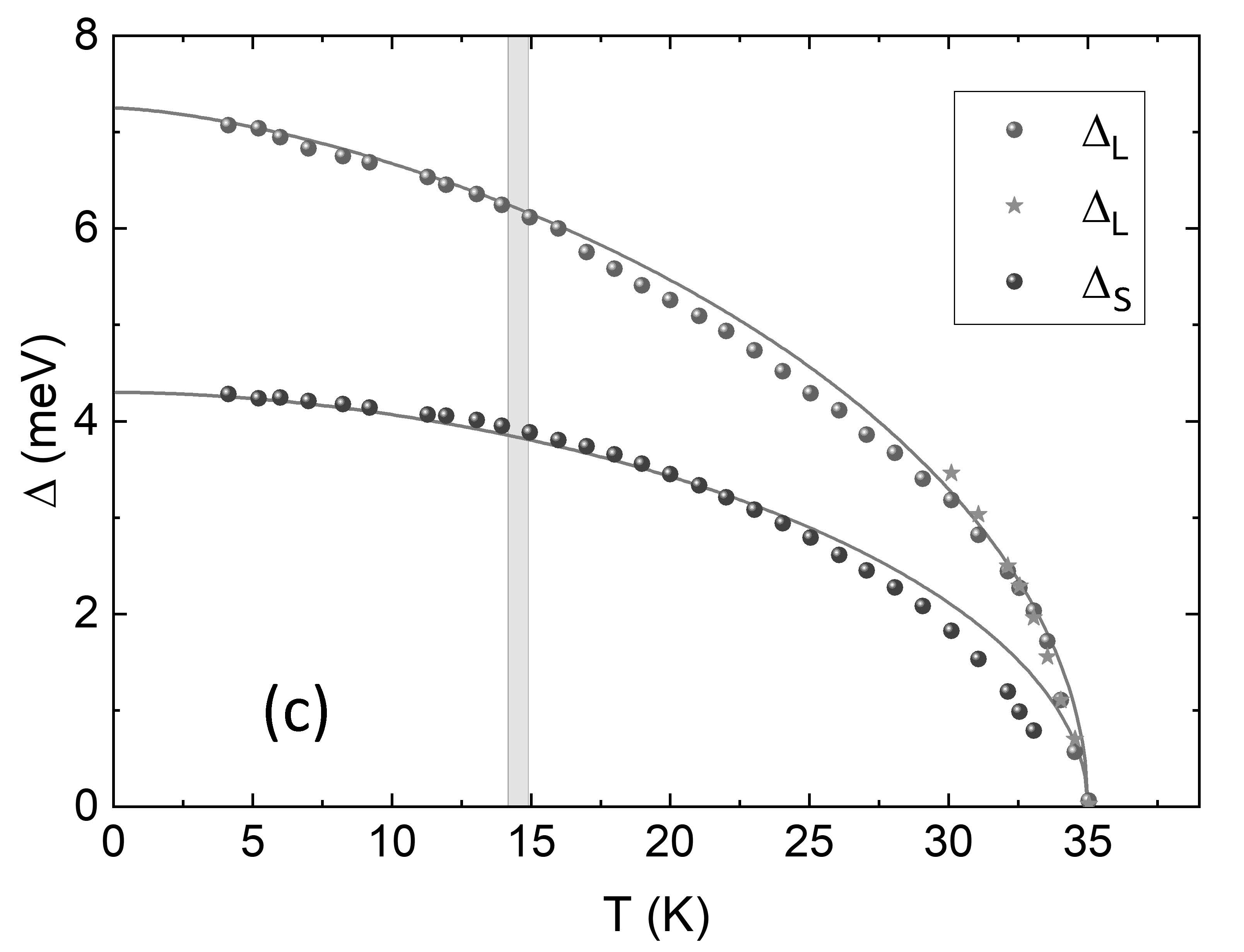}
\caption{(a) Typical $I - V$ characteristics (right $Y $ axis), and dynamic conductivity (left $Y$ axis) for two different junctions: the single and double-stack one. In the latter case, the  voltage is scaled down by
  $m = 2$ times. The shaded vertical stripes highlight positions of the
  $dI/dV$ minima. (b) 3D picture of a complete set of $dI/dV$ curves versus bias voltage $V$  and $T$ for the double-stack junction  $m=2$.
$V_i$ values are doubled here  as compared with the top panel.
(c) Temperature dependence of the two strongest gap features obtained from the  $dI/dV$ minima. Green circles represent the large gap, blue ones - the small gap. The red asterisks illustrate the reproducibility of the data in various measurements.
Solid gray lines represent the BCS-like fit. Panels (a) and (c) are quoted from \cite{kim_PRB_2021}.}
\label{fig:I-V_dI-dV}
\end{figure}

On the current-voltage characteristic (Fig.~\ref{fig:I-V_dI-dV})a the enhanced
conductivity is seen in the range $\pm 10$\,mV, typical for the Andreev reflection regime. On the dynamic
conductivity $ dI/dV$ curves (Fig.~\ref{fig:I-V_dI-dV}, top) one can see minima at bias voltages
$ V = \pm 8.6 $ and $ \pm 14.5$\,meV, indicating the presence of two superconducting condensates
with a ``large''
$\Delta_L $ and a ``small'' $\Delta_S$ superconducting gaps, respectively. As temperature rises,
the positions of the minima shift towards lower voltages and converge to zero at the same value
$ T_c \approx 35$\,K. This value is consistent with the bulk $ T_c $ determined from simultaneous measurements
of conductivity, as well as from thermodynamic measurements of susceptibility and magnetization, carried out on
samples from the same batch.

General  evolution of the gap amplitudes  with temperature is shown on   Fig.~\ref{fig:I-V_dI-dV}b.
It clearly shows the presence of at least two gaps in the spectrum,   $\Delta_L$ and $\Delta_S$.
Positions of the two major minima in $ dI/dV $ as a function  of temperature
are given in Fig.~\ref{fig:I-V_dI-dV}c.
By extrapolating the $\Delta_i(T)$ values to $T = 0$,   using the
BCS-type dependence,
we obtain $\Delta_L(0)=7.25\pm 0.3$\,meV and $\Delta_S(0)=4.3\pm 0.2$\,meV.
With these gap values, the characteristic ratios are $2\Delta_L/k_B T_c=4.80$ and $2\Delta_S/k_BT_c=2.85$.

From the data of Fig.~\ref{fig:I-V_dI-dV}c one can draw an important conclusion:
both dependencies, $\Delta_{L,S}(T)$ develop gradually with no features through
the temperature  $T_m\approx 15$\,K of Eu magnetic ordering. The  absence of features  in
$\Delta_i(T)$ in the vicinity of magnetic ordering  $T_m$, as well as the gap values agree with the ARPES data.
Thus, both Andreev reflection spectroscopy and ARPES clearly indicate that the amplitude of the superconducting
order parameter in RbEuFe$_4$As$_4$ is insensitive to magnetic ordering of the sublattice of  Eu atoms.

From high resolution ARPES measurements in Ref.~\cite{kim_PRB_2021}
the opening of a superconducting gap was observed in the electron and hole regions of the FS at
temperatures below $T_c$.
Since ARPES measurements are momentum resolved, the gap values $\Delta_i$ may be tied to the FS pockets:
the largest gap opens on the FS inner hole cylinder, as well as in other iron-pnictides
of the  122 family \cite{ding_EPL_2008, evtushinsky_PRB_2009, evtushinsky_PRB_2014}.
Within the measurement uncertainty, the same value $\approx 9$\,meV has a gap at the electron pocket.
In general, the gap values $\Delta_i$ determined from ARPES are in agreement with those determined from BJ measurements.
We stress that there are no singularities at temperatures near $T_m $ on the temperature dependence of the gap obtained from ARPES \cite{kim_PRB_2021}, as well as on the dependence from BJ measurements
(see Fig.~\ref{fig:I-V_dI-dV})c.

Local measurements by the scanning tunneling spectroscopy (STS)  also revealed a superconducting gap
\cite{stolyarov_JPCL_2020}, however its quantitative measurements were impeded by the fact that the cleaved
crystal surface of EuRb-1144 corresponds to the ``non-superconducting planes'' of Eu- and Rb atoms located far
from the FeAs ``superconducting planes'' [see Fig.~\ref{fig:structure}(a)].

\subsection{Band structure, Fermi surface and  Eu energy levels
from  ARPES, STS measurements, and from calculations}
\begin{figure}
\includegraphics[width=230pt]{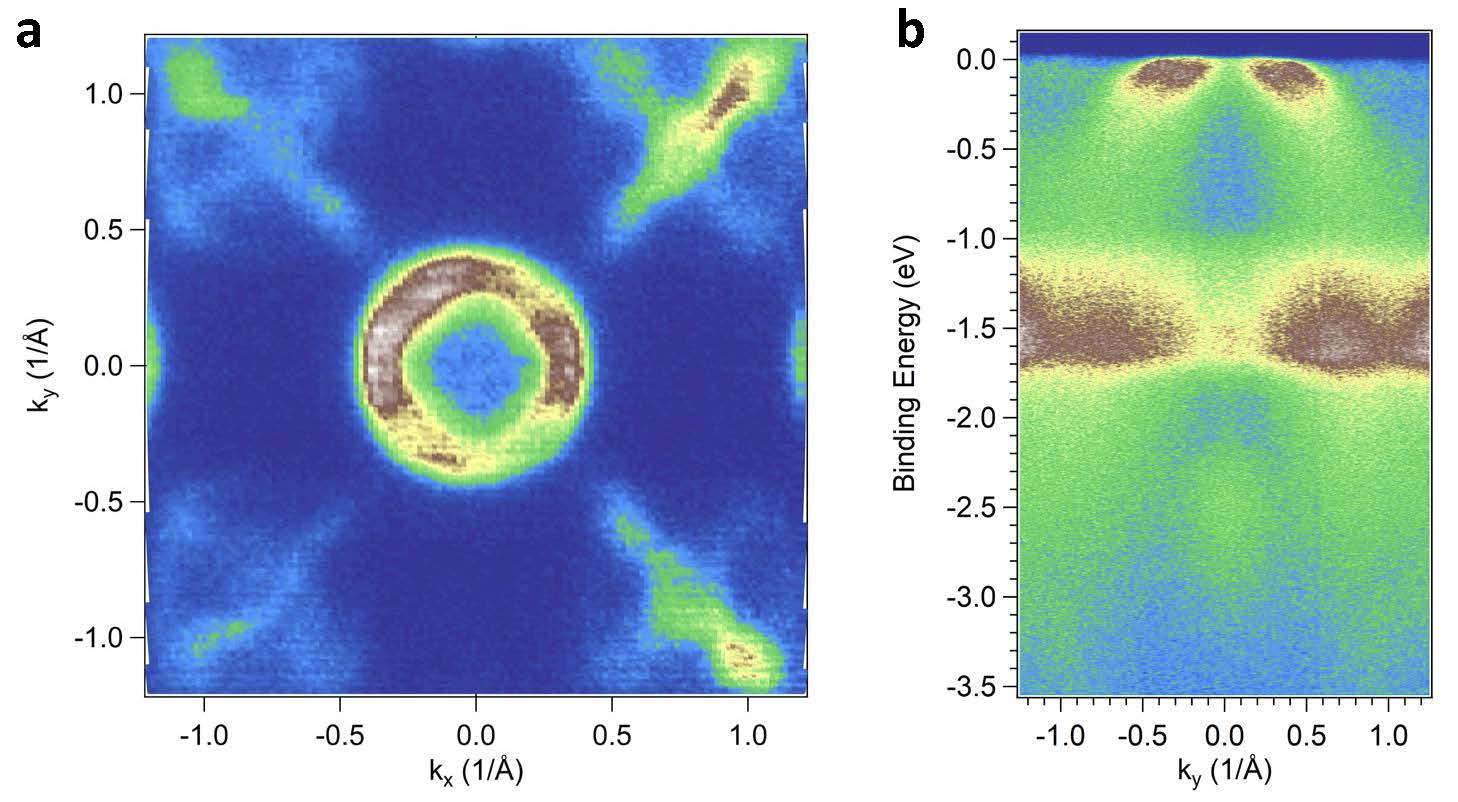}
   \includegraphics[width=115pt, height=130pt]{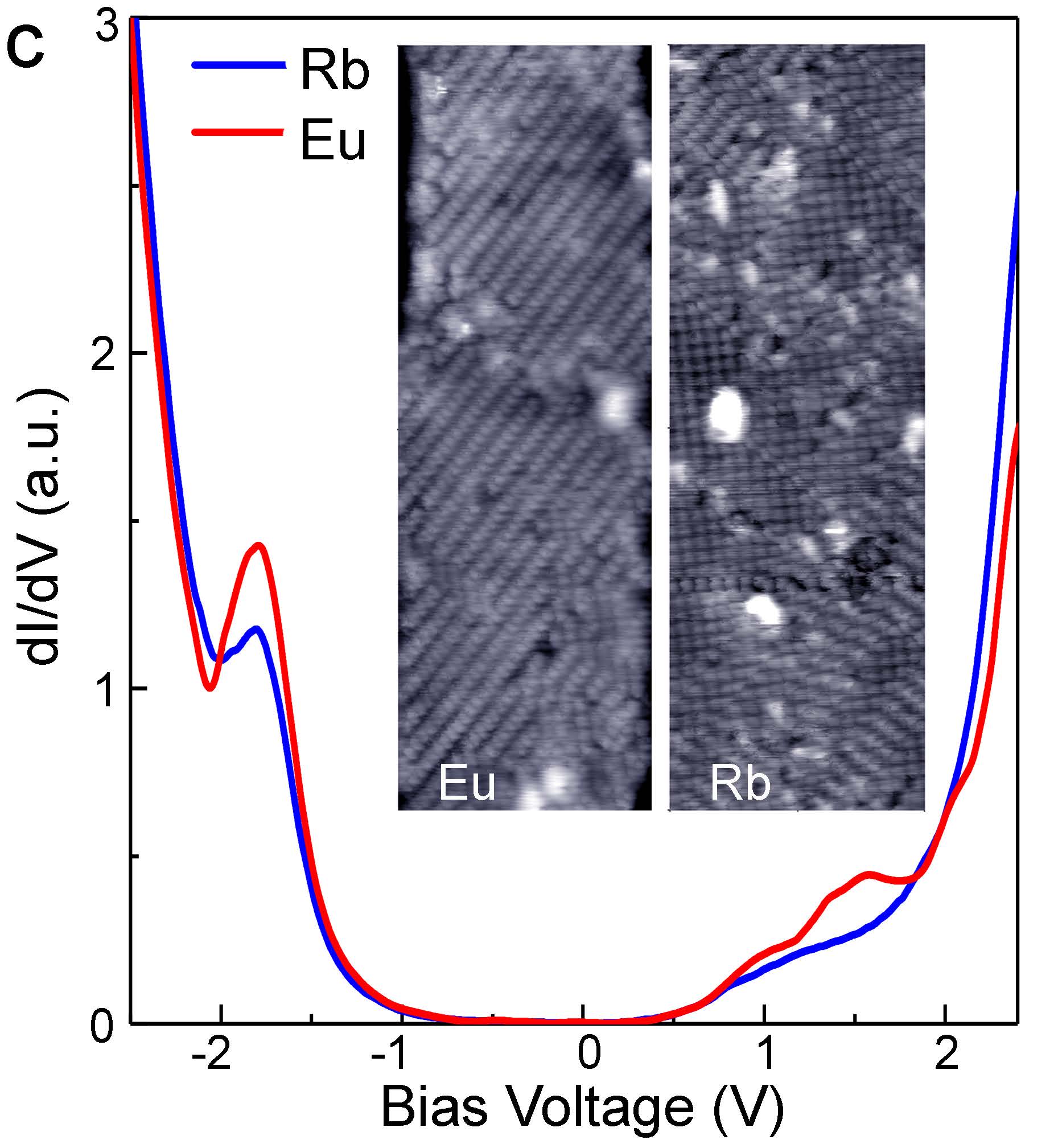}
   \includegraphics[width=100pt]{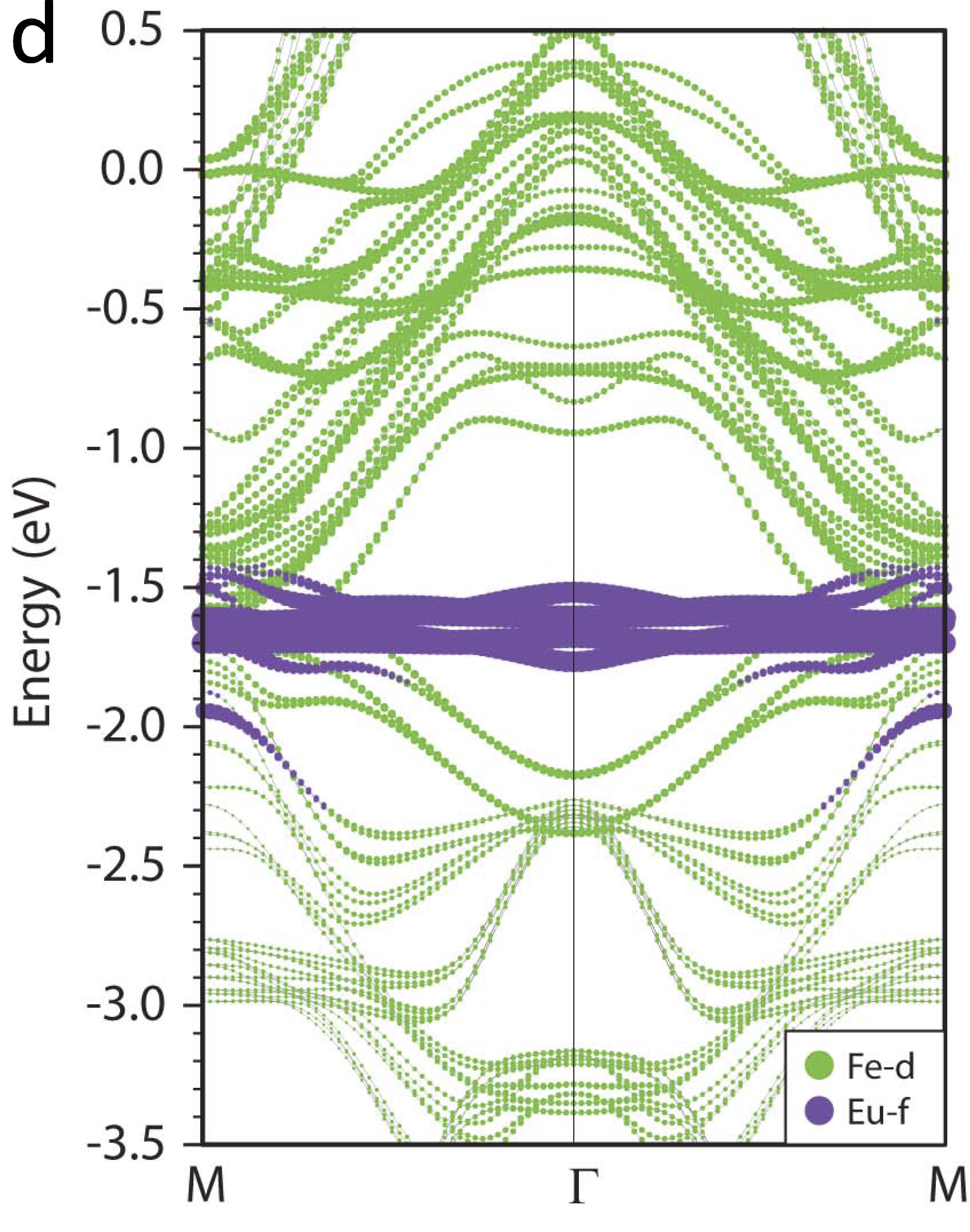}
\caption{(a) Map of the FS cross section by the $k_z = 0$ plane, measured at $T=40$\,K with a photon energy of 70\,eV. (b) Dispersion in the M - $\Gamma$ - M direction measured with a photon energy of 113\,eV. Brown corresponds to the photoemission intensity maximum, dark blue - to the intensity minimum. (c) Tunneling conductivity $dI/dV$ versus $V$ determined two different terminations, Eu and Rb, respectively. The insets show  $14\times 40$\,nm$^2$ areas, where the averaged spectra were obtained. (d) Calculated band dispersion in the
  M - $\Gamma$ - M direction: Fe-3d bands are shown with green lines, Eu-4f bands - violet. Adapted from \cite{kim_PRB_2021}.
}
\label{fig:FS-topology}
\end{figure}

Figure~\ref{fig:FS-topology}(a) shows a Fermi surface (FS) map in the $ k_x - k_y $ plane (for $k_z = 0$,
i.e. around $\Gamma $ point),
measured at a photon energy of 70\,eV. It can be seen  that the FS for EuRb-1144 is qualitatively similar to
FS of optimally doped Ba$_{1-x}$K$_x$Fe$_2$As$_2$ (the so-called ``Ba-122 '') ferropnictides
\cite{evtushinsky_PRB_2009, evtushinsky_PRB_2014}
and consists of three nested hole cylinders in the center of the Brillouin zone,
and more complex electronic pockets in the corners of the Brillouin zone.

The band dispersion measured in the
 M - $\Gamma$ - M  direction [Fig.~\ref{fig:FS-topology}(b)] besides the branches originating mainly from 3d Fe orbitals,
also includes branches from the Eu 4f orbitals, ranging from 1.0 to
1.7\,eV below the Fermi level. This ARPES result is consistent with the band structure calculations given in
in Fig.~\ref{fig:FS-topology}(d), where a bunch of almost flat Eu-zones  is seen in the region around -1.7\,eV. Thus, the spectral density of Eu 4f states  lies far away in energy from the
electronic states of Fe near the Fermi level, which are the states involved in superconducting pairing. For this reason, one also cannot expect a noticeable contribution of the Eu states
into superconducting pairing.
The measured Fermi surface cross sections and the band structure are in good agreement with the results of DFT calculations (see Fig.~\ref{fig:FS-topology}(d) and also Ref.~\cite{kim_PRB_2021}).

The described above results of ARPES measurements of the energy levels for Eu atoms  are confirmed by measurements of the tunneling conductivity. Typical tunneling conductivity spectrum measured by scanning tunneling spectroscopy
(STS) is shown in Fig.~\ref{fig:FS-topology}(c).
In this figure, a resonance is visible at bias voltage $ V\approx -1.8$\,V. The position of this resonance is consistent with the ARPES results [Fig.~\ref{fig:FS-topology}(b)] and DFT calculations [Fig.~\ref{fig:FS-topology}(d)].

\subsection{Eu magnetic levels
from  ResPES measurements and calculated magnetic structure of the Eu sublattice}

\begin{figure}[h]
   \includegraphics[width=0.5\textwidth]{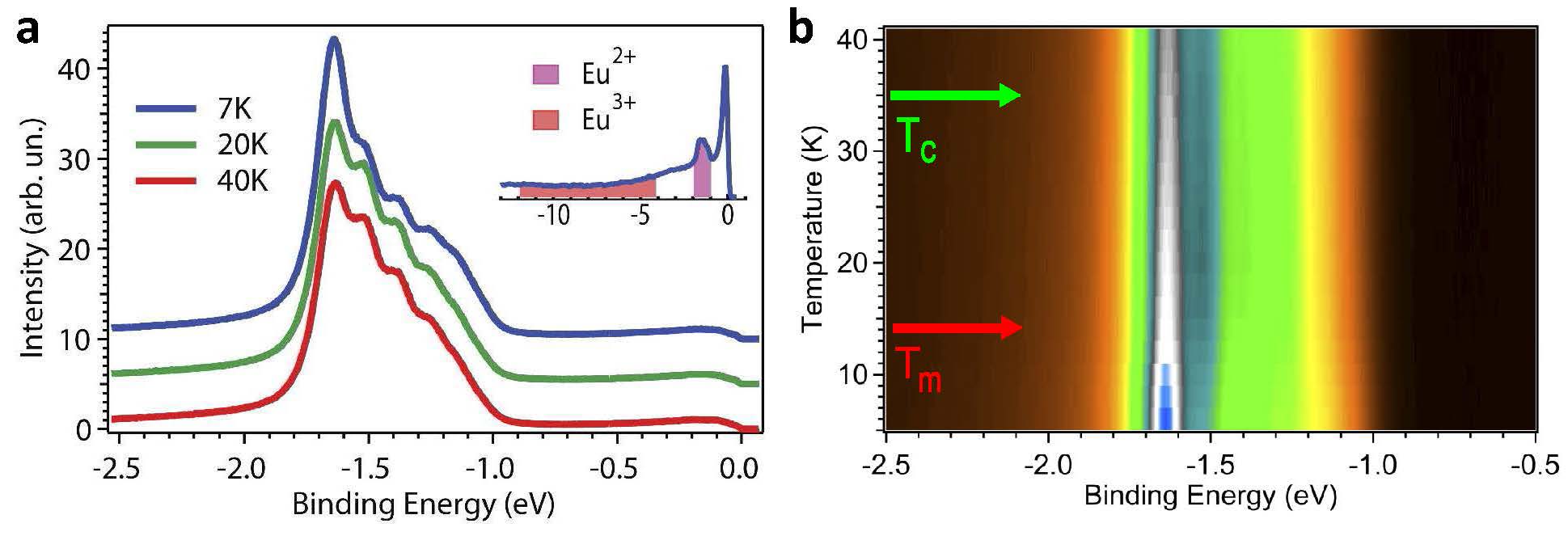}
\caption{
(a) Resonant photoemission spectra of Eu 4d $\rightarrow$ 4f
measured at temperatures of 7, 20 and 40\,K. The off-resonant 120\,eV survey
scan shown as insert  indicates no admixture of trivalent Eu state.
(b) Two-dimensional color presentation of
the intensity evolution of the Eu$^{2+}$ 4f resonant photoemission signal  in the temperature range from 40\,K to 5\,K, where white(blue) corresponds to the photoemission intensity maximum, whereas black  - to the intensity minimum. The two arrows mark the temperature of magnetic ordering and superconducting transition. Adapted from Ref.~\cite{kim_PRB_2021}.
}
\label{fig:ResPES}
\end{figure}

The fine structure of the Eu  levels was measured in Ref.~\cite{kim_PRB_2021} by
resonance photoemission spectroscopy at the absorption threshold of Eu 4d $\rightarrow$ 4f using photons with an energy of 142\,eV. The measurements were carried out by systematically varying temperature  from 40\,K to 5\,K,
i.e. in the normal state, via the superconducting state with a magnetically disordered Eu sublattice ($35> T> 15$\,K) into the magnetically ordered superconducting state ($T <14 $\,K); the results are shown in Fig.~\ref{fig:ResPES}.
The divalent  Eu state with 4f$^7$ electron configuration has a large and purely spin magnetic moment $ 7\mu_B$, which determines the complex magnetic properties of the Eu sublattice.

 As can be seen from Fig.~\ref{fig:ResPES}, when the temperature drops well below $T_m \approx 14 $\,K, the spectral weight is transferred from the side resonances to the central peak corresponding to the ground state. This  can be explained as the appearance of the long-range FM order of the 4f moments in the Eu layer. When the FM order sets in due to exchange interaction, which couples neighboring 4f moments, the angular momenta assume long-range orientation leading to a preferred direction between incoming light and angular momentum. The latter implies a different excitation probability for the ferromagnetically ordered 4f states than for the paramagnetically disordered 4f states with nearly isotropic angular momentum orientation. The inset in this figure shows a wide ``tail''
 of the valence band spectrum with no spectral structure in the binding energy range 5-8\,eV, which would indicate the presence of trivalent Eu. This clearly proves that the Eu atoms in RbEuFe $_4$As$_4$ are in the Eu $^{2+}$ state, without an admixture of the Eu$^{3+}$ states.
 The magnetic moment of Eu $6.97\mu_B$ calculated within the DFT also agrees well with the experiment.

Electronic spectra in Ref.~\cite{kim_PRB_2021} have been calculated within DFT for three different  types of Eu magnetic moment ordering:
FM, AFM-180 and AFM-90, matching fully ferromagnetic ordering and helical AFM
ordering with magnetization vector rotating between adjacent Eu layers by an angle of $ 180^\circ$ and
$90^\circ$, respectively. The latter  configuration was theoretically suggested in Ref.~\cite{koshelev-helical_PRB_2019} and reported experimentally from Bragg's
neutron scattering \cite{iida_PRB_2019}. The calculations in Ref.~\cite{kim_PRB_2021} showed that, regardless of the type of magnetic ordering of Eu,  the bands forming electronic M-pockets at the Fermi level are almost completely determined
by the  $d_{xz(yz)}$ orbitals of Fe, whereas the bands forming the hole $\Gamma$-pockets are determined by a mixture of different Fe 3d
orbitals. The main result of these calculations is that the band structure
 at the Fermi level and the orbital character of Fe 3d bands  are almost unaffected by the type of magnetic ordering
of Eu 4f magnetic moments. For this reason, the pairing interaction among Fe 3d itinerant electrons
is insensitive to the onset of the local magnetic order of Eu atoms.

By comparing the total energies of three different configurations of the Eu  magnetic ordering,  it was found in
Ref.~\cite{kim_PRB_2021}  that the AFM-90 configuration is energetically most favorable. The total energy of the AFM-180 is at 1.8 meV/f.u. higher, whereas the FM configuration is the least favorable being of 4.4\,meV/f.u. higher in energy than AFM-90 \cite{kim_PRB_2021}.

\section{Conclusion}
In the last three years, the recently discovered magnetic superconductor of the stoichiometric
composition RbEuFe$_4$As$_4$, has been the focus of intensive experimental  research,  including
high resolution ARPES and ResPES, STS, Andreev spectroscopy, electron spin resonance (ESR), infrared spectroscopy,
and measurements of thermodynamic properties.
As a result of these comprehensive experimental studies and theoretical ab-initio DFT calculations,
the helicoidal type of magnetic ordering of the  Eu moments was confirmed; it was theoretically established that the lowest energy is intrinsic to the helicoidal type AFM-90 of magnetic structure with
magnetization vector rotation by $90^\circ$ between the Eu layers.

ARPES measurements showed that the band structure of EuRb-1144 is
very similar to the structure of parent compounds of the 122 family of iron-based superconducting non-magnetic
pnictides. The band structure measured by  ARPES and confirmed by STS is in agreement with the theoretically calculated one. These results clearly show the position of Eu levels with a high spectral density at
$\sim 1.7$\,eV  below the Fermi energy. The DFT calculations also indicate that the type of magnetic ordering
is weakly reflected in the electronic spectrum of the 3d itinerant Fe states near the Fermi level, which participate in the superconducting pairing.

Measurements of superconducting properties using thermodynamic techniques, ARPES and Andreev spectroscopy
have identified fully gapped state with the $s$-type symmetry of the order parameter, slightly anisotropic in the $k_x-k_y$ plane. Temperature dependencies of the order parameter amplitude for different pockets of the FS are consistent with the BCS model. In this way,  the superconducting properties of EuRb-1144, in general, turn out to be almost  insensitive to the Eu- magnetic ordering itself and to its type.

However, one cannot extend the results described here too broadly and to  assume that superconductivity
and magnetic ordering are completely independent. In recent work \cite{koshelev_PRB_2020} it was predicted
that the magnetic ordering at $T = T_m$ should  affect the density of the superconducting condensate $\rho_s$, rather than  the amplitude of the superconducting order parameter. First measurements of the spatial profile of the Abrikosov vortices \cite{collomb_PRL_2021} confirm this theoretical prediction. An anomaly in specific heat at $T = 15 $ K found in Ref.~\cite{willa_PRB_2019} were interpreted by the authors as the Berezinsky-Kosterlitz-Thouless (BKT) transition. The temperature evolution of the ESR line was also discussed in terms of the BKT topological transition
\cite{hemmida_arxive}.

Another interesting question is the mechanism for establishing the long-range magnetic order of the Eu layers: since the Eu-Eu spacing is too large (1.3\,nm), the itinerant electrons of the Fe orbitals simultaneously participate in s-type pairing and transfer the spin moment
between the Eu layers. One  cannot also rule out the existence of phase transitions between
various types of magnetic ordering, including more complex ones than the helix.

\section{Acknowledgments }
Crystal synthesis, characterization, magnetotransport, and Andreev spectroscopy were performed
using the equipment of the LPI Shared facility center. A.S.U., A.V.S., and V.A.V. are grateful for financial support within the framework of the State assignment  (project ``Physics of high-temperature superconductors and
new quantum materials'' No. 0023-2019-0005). K.S.P. and V.M.P. are grateful for the support of the Russian Science Foundation (project No. 21-12-00394).
S.V.E. is grateful for financial support within the framework of the State assignment  of IPPM SB RAS (project FWRW-2019-0032).

\end{document}